\documentstyle[12pt]{article}
\input{psfig}
\catcode`@=11
\def\seceqaa{\@addtoreset{equation}{section}
           \def\theequation{A\arabic{equation}}}
\def\seceqbb{\@addtoreset{equation}{section}
           \def\theequation{B\arabic{equation}}}
\def\seceqcc{\@addtoreset{equation}{section}
           \def\theequation{C\arabic{equation}}}
\catcode`@=12

\begin{document}

\title{Correct Treatment of ${1\over{(\eta\cdot k)^p}}$
Singularities in the Axial Gauge Propagator}
\author{{Satish. D. Joglekar}\thanks{e-mail:sdj@iitk.ac.in},
{A. Misra} \thanks{e-mail:aalok@iitk.ac.in}\\
Department of Physics, Indian Institute of Technology,\\
 Kanpur 208 016, UP, India}
\maketitle

\begin{abstract}
The propagators in axial-type, light-cone and planar gauges contain
${1\over{(\eta\cdot k)^p}}$-type singularities. These
singularities have generally been treated
by inventing prescriptions for them. 
In this work, we propose an alternative procedure
for treating these singularities in the path integral formalism using
the known way of treating the singularities
in Lorentz gauges. To this end, we use a finite field-dependent
BRS transformation that interpolates between
Lorentz-type and the axial-type gauges. We arrive at
the $\epsilon$-dependent tree propagator in the axial-type gauges.
We examine the singularity structure of  the propagator and
find that the axial gauge propagator so constructed has
{\it no} spurious poles (for real $k$). It however has
a complicated structure in a small region near $\eta\cdot k=0$.
We show how this complicated structure can
effectively be replaced by a much simpler propagator.
\end{abstract}

\section{Introduction}

As far as we know today, the known high energy physics is well explained
by the Standard Model (SM).
SM is an SU(3)$\times$SU(2)$\times$U(1) nonabelian gauge theory  \cite{cl}. 
Hence, the practical calculations in electroweak and strong
interactions are calculations in a gauge theory
requiring a choice of gauge. The two choices of gauges most 
frequently employed are the Lorentz-type
and the axial-type gauges. (The latter include the light-cone gauges (LCG)
and planar gauges, while the former include $R_\xi$-gauges in spontaneously
broken gauge theories (SBGT).) The Lorentz-type gauges have been
popular on account of their Lorentz covariance, simplicity of
Feynman rules and availability of a gauge parameter to ensure
gauge independence of physical results. The disadvantage of 
Lorentz-type gauges is however the presence of Faddeev-Popov
ghosts and relatively large number of Feynman diagrams
needed for evaluation of Green's functions.
The axial-type gauges, on the other hand, have the advantage
of formal decoupling of ghosts \cite{diff}. This leads to a much  smaller
number of diagrams to be  evaluated. These gauges are particularly useful 
in perturbative QCD calculations \cite{bass1}. In fact the first
QCD calculations  were done in these gauges \cite{gw}.

The main disadvantage of axial-type gauges arises from the lack of Lorentz
covariance and especially from the appearance of $1/(\eta\cdot k)^p$-type
spurious singularities in propagators.  Much literature
has been devoted  to the question of how these singularities should
be treated \cite{diff,bass2}. Prescriptions
have been proposed to deal with this issue:
two important ones of these are the ``principal-value prescription "
(PVP)\cite{pvp} and the Mandelstam-Leibbrandt (ML) prescriptions 
\cite{lm}. These, however, have lead in many cases to difficulties. The PVP
procedure fails for LCG already at the on loop level and yields the wrong
answer for the Wilson loop to order $g^4$\cite{diff}.

Moreover, there are instances where
the ghosts need to be taken into account \cite{diff}. In canonical
quantization, the treatment of $1/\eta\cdot k$-type singularities has
been given for axial gauges of the from $A_1+\lambda A_3=0$\cite{landsh}
(This does not include LCG).

In this work, we advocate an ab-initio and fresh
approach to the question of the axial gauge propagator
based on earlier works \cite{jm,jb,talk,BRSpap1,BRS2lett}. The approach
here utilizes a finite field-dependent BRS transformation established earlier
between the Lorentz-type and the axial-type gauge\cite{jm,jb,talk}. This
transformation has been used to write down a compact expression that interpolates
between Green's functions from the 
axial-type gauges to the Lorentz-type gauges \cite{BRSpap1}.
We apply the results of \cite{BRSpap1} to the relation between the
axial and the Lorentz gauge propagator. The procedure we adopt is detailed
below.

We know how the $1/k^2$ singularities of the propagator are handled in the
Lorentz gauges. These are in effect,
treated by replacing $k^2$ by $k^2+i\epsilon$. Where the poles have physical
interpretation, this amount to propagation of positive frequency waves into
future and negative frequency waves into
the past as the Feynman propagator for a physical field shows. This
is taken into account in a Lorentz covariant
manner by introduction of a term $\epsilon\int d^4x(A^2/2-{\bar c}c)$ to
the action. Introduction of such a term also has
natural interpretation in the Minkowskian formulation of Lorentz gauge theories.
This is elaborated in Section 3. We start from this well-established
procedure in Lorentz gauges. We then perform a finite field-dependent 
BRS transformation \cite{jm,jb} (a nonlocal field transformation) that
converts the Lorentz to the axial gauges. This procedure,
following the work \cite{BRS2lett} leads us to an expression 
for the tree $\epsilon$-dependent
propagator for the gauge field in the axial gauges. We suggest 
that this expression, should, in principle be used for
the axial  gauges propagator. We analyze
the singularity structure of the propagator and find that for
real $\eta\cdot k$, there are no spurious
poles. The propagator reduces to the usual
propagator for $|\eta\cdot k|>>\epsilon$;
however, it show a complex structure in a small region near
$\eta\cdot k =0$. We show that this propagator can equally
well be replaced by an effective simpler expression. 

Our prescription, by its very construction, has the
desirable propertry that it preserves the value of the Wilson
loop \cite{CT}. This is so because the field transformation of \cite{jm} and
\cite{BRSpap1} was explicitly constructed to preserve the expectation values of
gauge-invariant observables as you go from gauge to gauge. This is
unlike the other prescriptions
where the property has to be imposed as a check on the prescription. For
more comments, please see reference \cite{inprepar}.

We now briefly state the plan of the paper. In Section 2,
we review the results needed in this work. In particular
we introduce the FFRBS transformations and the results of 
references \cite{BRSpap1} and \cite{BRS2lett}. In Section 3, we obtain
the effect
of $\epsilon\int d^4x(A^2/2-{\bar c}c)$ term in the Lorentz
gauge generating functional. In Section 4, we
work out the $\epsilon$-dependent axial gauge propagator in detail.
In Section 5, we examine the singularity structure of 
the propagator. In Section 6, we show how the propagator can effectively be 
replaced by a much simpler expression.
Section 7 has the conclusion.

\section{Summary of Results on FFBRS Transformation  between Lorentz- 
and Axial-type Gauges}

\subsection{Notations and Conventions}

We start with the Faddeev-Popov effective action (FPEA) in linear 
Lorentz-type gauges:
\begin{equation}
\label{eq:FPEAL}
S^L_{\rm eff}[A, c, {\bar c}]=\int d^4x\biggl(-{1\over 4} F^\alpha_{\mu\nu}
F^{\alpha, \mu\nu}\biggr)+S_{\rm gf}+S_{\rm gh},
\end{equation}
where the gauge-fixing
action $S_{\rm gf}$ is given by:
\begin{equation}
\label{eq:gfL}
S^L_{\rm gf}=-{1\over{2\lambda}}
\int d^4x\sum_{\alpha}(\partial\cdot A^\alpha)^2\equiv-{1\over{2\lambda}}
\int d^4x\sum_{\alpha}(f^\alpha_L[A])^2,
\end{equation}
and the ghost action $S_{\rm gh}$ is given by:
\begin{equation}
\label{eq:ghL}
S^L_{\rm gh}=-\int d^4x {\bar c}^\alpha M^{\alpha\beta}c^\beta,
\end{equation}
where
\begin{equation}
\label{eq:MLdef}
M^{\alpha\beta}[A(x)] \equiv \partial^\mu D^{\alpha\beta}_\mu(A,x).
\end{equation}
The covariant derivative is defined by:
\begin{equation}
\label{eq:covDdef}
{\rm D}^{\alpha\beta}_\mu\equiv\delta^{\alpha\beta}\partial_\mu
+gf^{\alpha\beta\gamma}A_\mu^\gamma.
\end{equation}

In a similar manner, the FPEA in axial-type gauges, is given by:
\begin{equation}
\label{eq:FPEAA}
S^A_{\rm gf}\equiv-{1\over{2\lambda}}\int d^4x\sum_{\alpha}
(\eta\cdot A^\alpha)^2\equiv-{1\over{2\lambda}}\sum_{\alpha}\int d^4x 
(f_A^\alpha[A])^2.
\end{equation}
We require $\eta_\mu$ to be real, but otherwise unrestricted.
and
\begin{equation}
\label{eq:ghA}
S^A_{\rm gh}=-\int d^4x {\bar c}^\alpha\tilde{M}^{\alpha\beta}c^\beta,
\end{equation}
with
\begin{equation}
\label{eq:tildeMdef}
\tilde{M}^{\alpha\beta}=\eta^\mu D^{\alpha\beta}_\mu.
\end{equation}
In the $\lambda\rightarrow 0$, 
\begin{equation}
\label{eq:lambdazero}
e^{iS^A_{\rm gf}}\sim \prod_{\alpha,x}\delta
\biggl(\eta\cdot A^\alpha(x)\biggr).
\end{equation}
Thus, in the presence of the delta function, the $A$-dependent term in 
$\tilde{M}$ can be dropped leading to the formally ghost-free matrix.
As is well known, $S^L_{\rm eff}$ and $S^A_{\rm eff}$ 
are invariant under the BRS transformations:
\begin{eqnarray}
\label{eq:BRS1}
& & \delta
 A^\alpha_\mu(x)=D_\mu^{\alpha\beta}c^\beta(x)\delta\Lambda
\nonumber\\
& & \delta c^\alpha(x)=-{g\over 2}f^{\alpha\beta\gamma}c^\beta(x)
c^\gamma(x)\delta \Lambda
\nonumber\\
& & \delta{\bar c}^\alpha(x)={{f^\alpha[A]}\over{\lambda}}\delta\Lambda,
\end{eqnarray}
where $f^\alpha[A]=\partial\cdot A^\alpha$ or $\eta\cdot A^\alpha$,
depending  on whether one  has action in the Lorentz or the axial-type gauges.
We also need the interpolating mixed gauge with
\begin{equation}
\label{eq:Mgf}
S^M_{\rm g.f.}=-{1\over{2\lambda}}\int d^4x
[\partial\cdot A^\alpha(1-\kappa)+\kappa\eta\cdot A^\alpha]^2
\end{equation}
and the associated ghost term
\begin{equation}
\label{eq:Mgh}
S^M_{\rm gh}=-\int d^4x{\bar c}\biggl[(1-\kappa)M+\kappa\tilde M\biggr]c.
\end{equation}
The net effective action $S^M_{\rm eff}$ 
has a BRS symmetry under transformations
(\ref{eq:BRS1}) with 
$f^\alpha[A]\to (1-\kappa)\partial\cdot A^\alpha+\kappa\eta\cdot A^\alpha$.

We write in general, the BRS transformations in this case as:
\begin{equation}
\label{eq:del12def}
\delta_{\rm BRS}\phi_i=(\tilde\delta_{1i}[\phi]
+\kappa\tilde\delta_{2i}[\phi])\delta\Lambda.
\end{equation}

\subsection{FFBRS Transformations}

As observed by Joglekar and Mandal \cite {jm}, in (\ref{eq:BRS1}),
$\delta\Lambda$ need not be infinitesimal
 nor need it be field-independent as long as it  does not  depend on
$x$ explicitly for (\ref{eq:BRS1}) to by a symmetry of FPEA
In fact, the following finite field-dependent BRS (FFBRS)
transformations were introduced:
\begin{eqnarray}
\label{eq:BRS2}
& & 
{{A^\prime}}\ ^\alpha_\mu = A^\alpha_\mu  
+ D_\mu^{\alpha\beta}c^\beta(x)\Theta[\phi]
\nonumber\\
& & {{c^\prime}}\ ^\alpha = c^\alpha-{g\over 2}f^{\alpha\beta\gamma}c^\beta(x)
c^\gamma(x)\Theta[\phi]
\nonumber\\
& & {{\bar c^\prime}}\ ^\alpha = {\bar c}^\alpha 
+ {{f^\alpha[A]}\over{\lambda}}\Theta[\phi],
\end{eqnarray}
or generically
\begin{equation}
\label{eq:BRS3}
\phi^\prime_i(x)=\phi_i(x)+\delta_{\rm BRS}\phi_i(x)\Theta[\phi],
\end{equation}
where $\Theta[\phi]$ is an $x$-independent functional of $A,\ c,\ {\bar c}$ (generically denoted
by $\phi_i$) and these were also the symmetry of the FPEA.  
The transformations of the  form ({\ref{eq:BRS2}) were used to
connect actions of different kinds for Yang-Mills theory in \cite {jm} and 
\cite {jb}. The FPEA is invariant under (\ref{eq:BRS2}), but the
functional measure is not invariant under the
(nonlocal) transformations (\ref{eq:BRS2}). The Jacobian for the
FFBRS transformations can be expressed (in special
cases dealt with in \cite{jm,jb}) effectively as
$exp(iS_1)$ and this $S_1$ explains the difference between the two
effective actions. Such FFBRS transformations were constructed in 
\cite{jm},\cite{jb} by
integration of an infinitesimal field-dependent BRS
(IFBRS) transformation:
\begin{equation}
\label{eq:BRS4}
{d\phi_i(x,\kappa)\over{d\kappa}}
=\delta_{\rm BRS}[\phi(x,\kappa)]\Theta^\prime[\phi(x,\kappa)]
\end{equation}
The integration of (\ref{eq:BRS4}) from $\kappa=0$ to 1, leads to the FFBRS transformation
of (\ref{eq:BRS3}) 
with $\phi(\kappa=1)\equiv \phi^\prime$ and $\phi(\kappa=0)=\phi$.
Further $\Theta$ in (\ref{eq:BRS3}) was related to $\Theta^\prime$ by:
\begin{equation}
\label{eq:BRS5}
\Theta[\phi]=\Theta^\prime[\phi]{{exp[f[\phi]]-1}\over{f[\phi]}},
\end{equation}
where
\begin{equation}
\label{eq:BRS6}
f[\phi]=\sum_i\int d^4x{\delta\Theta^\prime\over{\delta\phi_i(x)}}\delta_{\rm BRS}\phi_i(x)
\end{equation}
FFBRS transformations of the type (\ref{eq:BRS3}) were used
to connect the FPEA in Lorentz-type gauges with  gauge parameter $\lambda$
to (i) the most general BRS/anti-BRS symmetric action in linear gauges,
(ii)FPEA in quadratic gauges, (iii) the FPEA in Lorentz-type gauges with another gauge parameter
$\lambda^\prime$ in \cite{jm}. It was also used to connect the former to FPEA
in axial-type gauges in \cite{jb}. We shall now summarize the results of \cite{jb}
in {\bf 2.3}. 

\subsection{ FFBRS Transformation for Lorentz to Axial Gauge $S_{\rm eff}$}

We give the results for the FFBRS transformation that connects 
the Lorentz-type gauges (See (\ref{eq:FPEAL})) with  gauges parameter $\lambda$ to 
axial gauges (See (\ref{eq:FPEAA})) with same gauge parameter $\lambda$.
[The same calculation can be used to connect it to axial gauges 
with another  gauge
parameter $\lambda^\prime$: one simply rescales $\eta$ suitably.] 
They are obtained by integrating:
\begin{equation}
\label{eq:BRS7}
{d\phi_i(\kappa)\over{d\kappa}}=\delta_{\rm BRS}[\phi]\Theta^\prime[\phi],
\end{equation}
with
\begin{equation}
\label{eq:Thetaprimedef}
\Theta^\prime=i\int d^4x{\bar c}^\alpha
(\partial\cdot A^\alpha-\eta\cdot A^\alpha).
\end{equation}
the consequent $\Theta[\phi]$ is given by (\ref{eq:BRS5}) with
\begin{equation}
\label{eq:fdef}
f[\phi]=i\int d^4x\Biggl[{\partial\cdot A^\alpha\over\lambda}
(\partial\cdot A^\alpha-\eta\cdot A^\alpha)
+{\bar c}(\partial\cdot{\rm D}-\eta\cdot{\rm D})c^\alpha\Biggr].
\end{equation}

The meaning of these field transformations is as follows.
Suppose we begin with vacuum expectation  value of  a gauge invariant functional
$G[\phi]$ in the Lorentz-type gauges:
\begin{equation}
\label{eq:vev}
\langle\langle G[\phi]\rangle\rangle\equiv\int{\cal D} 
\phi G[\phi]e^{iS^L_{\rm eff}[\phi]}.
\end{equation}
Now, we perform the transformation $\phi\rightarrow\phi^\prime$ given by (\ref{eq:BRS3}).
Then we have [with $G[\phi^\prime]=G[\phi]$ by gauge invariance]
\begin{equation}
\label{eq:vevFFBRS}
\langle\langle G[\phi]
\rangle\rangle\equiv\langle\langle G[\phi^\prime]\rangle\rangle
=\int {\cal D}\phi^\prime J[\phi^\prime]G[\phi^\prime]e^{iS_{\rm eff}^L[\phi^\prime]}
\end{equation}
on account of the BRS invariance of $S_{\rm eff}^L$. 
Here $J[\phi^\prime]$ is the Jacobian
\begin{equation}
\label{eq:Jdef}
{\cal D}\phi={\cal D}\phi^\prime J[\phi^\prime].
\end{equation}
As was shown in \cite{jm}, for the special case $G[\phi]\equiv{\bf 1}$,  the
Jacobian $J[\phi^\prime]$ in (\ref{eq:Jdef}), can be replaced by 
$e^{iS_1[\phi^\prime]}$ where
\begin{equation}
\label{eq:S_eff^Adef}
S^L_{\rm eff}[\phi^\prime]+S_1[\phi^\prime]=S^A_{\rm eff}[\phi^\prime].
\end{equation}
As shown in Section III of \cite{BRSpap1}, this
replacement is valid for  any gauge invariant $G[\phi]$ 
functional of  $A$. If one were to live  with vacuum expectation values of 
gauge invariant observables,  the FFBRS in  \cite{jm} would be
sufficient. But as  seen in \cite{BRSpap1},  general Green's functions need
a modified treatment.

\subsection{Relations between Green's  functions in Axial-
and Lorentz-type gauges}

The FFBRS in {\bf 2.3} was used to correlate arbitrary Green's
functions in  the Lorentz-type and Axial-type gauges
\cite{BRSpap1} Let $O[\phi]$ represent any field operator (local or
multi-local). Then the relation between the Green's functions  in the two
gauges is given by:
\begin{eqnarray}
\label{eq:relAL}
& & \langle\langle O[\phi]\rangle\rangle_A\equiv\int{\cal D}\phi^\prime 
O[\phi^\prime]e^{iS^A_{\rm eff}[\phi^\prime]}\nonumber\\
& & =\int {\cal D}\phi\biggl(O[\phi]+\sum_i\delta_i\phi[\phi]
{\delta O\over{\delta\phi_i}}\biggr)e^{iS^L_{\rm eff}[\phi]}\nonumber\\
& & \equiv\langle\langle O[\phi]+\sum_i\delta_i\phi[\phi]
{\delta O\over{\delta\phi_i}}\rangle\rangle_L,
\end{eqnarray}
where
\begin{eqnarray}
\label{eq:result1}
& & \phi^\prime=\phi+\biggl(\tilde\delta_1[\phi]\Theta_1[\phi]
+\tilde\delta_2[\phi]\Theta_2[\phi]\biggr)\Theta^\prime[\phi]
\nonumber\\
& & \equiv\phi+\delta\phi[\phi]
\end{eqnarray}
is an FFBRS \cite{jm} with
\begin{equation}
\label{eq:Theta12def}
\Theta_{1,2}[\phi]\equiv\int_0^1 d\kappa (1,\kappa)exp\biggl(\kappa
f_1[\phi]
+{\kappa^2\over 2}f_2[\phi]\biggr);
\end{equation}
\begin{eqnarray}
\label{eq:f12def2}   
& & f_1[\phi]\equiv i\int d^4x\biggl[{\partial\cdot A^\alpha\over\lambda}
(\partial\cdot A^\alpha-\eta\cdot A^\alpha)+{\bar c}
(\partial\cdot{\rm D}-\eta\cdot{\rm D})c\biggr]
\nonumber\\
& & f_2[\phi]\equiv -{i\over\lambda}
\int d^4x (\partial\cdot A^\alpha-\eta\cdot A^\alpha)^2,
\end{eqnarray}and
\begin{equation}
\label{Thprimedef}   
\Theta^\prime\equiv i\int d^4x {\bar c}^\alpha
(\partial\cdot A^\alpha-\eta\cdot A^\alpha).
\end{equation}
The relation (\ref{eq:relAL}) can be used to related the ordinary
Green's functions, operator Green's functions, 
etc. in the two set of gauges depending on
the choice of $O[\phi]$. 

A much more convenient and tractable result was
also derived in \cite{BRS2lett}
\begin{equation}
\label{eq:AtoL2}
\langle {\cal O}\rangle_A=\langle{\cal O}\rangle_L+\int_0^1 d\kappa\int
D\phi
\sum_i\biggl(\tilde\delta_{1,i}[\phi]+\kappa\tilde\delta_{2,i}[\phi]
\biggr)\Theta^\prime[\phi]
{\delta{\cal O}\over{\delta\phi_i}} e^{iS^M_{\rm eff}},
\end{equation}
where $S_{\rm eff}^M$ is the
FPEA  for the
mixed gauge function with the gauge fixing term
defined in (\ref{eq:Mgf}) and $\tilde\delta_{1,i}$ and
$\tilde\delta_{2,i}$ have been defined in 
(\ref{eq:del12def}).
Of course, $\tilde\delta_{2,i}$ is non-vanishing only for ${\bar c}$ field.

\section{The General Procedure for Generating Prescription}

In this section, we shall outline the general procedure for
generating the correct treatment for ${1\over{\eta\cdot k}}$ 
singularities in the axial-type gauges
starting from the Lorentz-type gauges. In the Lorentz-type
gauges, also there is a 
singularity in the propagator at $k^2=0$ in both the gauge and
the ghost propagators. This, in 
analogy with the scalar particle, is dealt with by adding
an $i\epsilon$ term, viz $k^2\rightarrow k^2+i\epsilon$
($\epsilon$ is small positive) in 
the denominators. As is well known, this 
prescription allows the propagation of positive
energy solutions into future and the negative energy 
solutions into the past. The role of this prescription 
in the Lorentz-type
gauges can also be understood clearly 
in the Minkowskian formulation of quantum field theory. 
The above prescription is implemented
by an addition of the term $-i{\epsilon\over 2}A_\mu A^\mu$
to the gauge field action. 
In the context of a scalar theory, the $i\epsilon\phi^2$ term provides a damping
in the path integral:
\begin{equation}
\label{eq:scleps}
W=\int{\cal D}\phi e^{i(S+i\epsilon\phi^2)}
\end{equation}
for large $\phi$. In the context of gauge theories, we expect the 
$\epsilon$-dependent term to be determined by similar damping
provided in the transverse degrees of freedom. Then the form of the term in
the context of a covariant formulation viz $-i\epsilon A_\mu A^\mu/2$ is
determined by covariance.
Thus, this treatment of the ${1\over k^2}$-type singularity is well 
understood in the Lorentz-type gauges. Further, 
there are WT identities for Green's 
functions that have terms that involve both the
ghost and the gauge propagators.
Their exact preservation requires that a similar 
modification be made in
the ghost propagator poles $1/k^2\rightarrow 1/{k^2+i\epsilon}$.
Thus, the path integral for $\langle 0|0\rangle$ in the Lorentz-
type gauges we normally start with, is given in Minkowski
space, by:
\begin{eqnarray} 
\label{eq:presp2}
& & W^L=\int {\cal D}\phi e^{i[S^L_{\rm eff}-i{\epsilon\over 2}
A_\mu A^\mu+i\epsilon {\bar c}c]}\nonumber\\
& & \equiv\int {\cal D}\phi
e^{iS^L_{\rm eff}[\phi]+iO_1[A,c,{\bar c},\epsilon]}.
\end{eqnarray}
We now expect that if we start with this $W^L$ that has no pole
prescription ambiguities and make suitable field transformation
(as outlined in Section 2) to the axial gauges, we
should obtain an ambiguity-free treatment of the axial gauge
propagator. We thus imagine making the field transformation
of (\ref{eq:BRS3}) viz:
\begin{equation}
\label{eq:presp3}
\phi^\prime_i(x)=\phi_i(x)+\delta_{\rm BRS}[\phi_i]\Theta[\phi]
\end{equation}
with $\Theta[\phi]$ given explicitly by (\ref{eq:BRS5}), (\ref{eq:Thetaprimedef})
and (\ref{eq:fdef}).

As shown in \cite{BRSpap1}, the effect of this field transformation can be 
evaluated via the formula (\ref{eq:relAL}), and is given by:
\begin{eqnarray} 
\label{eq:presp4}
& & W^L=\int{\cal D}\phi e^{iS^L_{\rm eff}[\phi]+i\epsilon O_1[\phi]}
\nonumber\\
& & =\int {\cal D}\phi^\prime e^{iS^A_{\rm eff}[\phi^\prime]+i\epsilon 
O_1^\prime[\phi^\prime]}
\end{eqnarray}
with
\begin{equation}
\label{eq:presp5}
O_1^\prime=O_1+\sum_i\delta\phi_i[\phi]{\delta O\over{\delta\phi_i}}.
\end{equation}
We regard the net exponent, including the $new$ ${\cal O}(\epsilon)$
terms, viz.
\begin{equation}
\label{eq:presp6}
{S^A_{\rm eff}}\ ^\prime\equiv S^A_{\rm eff}+\epsilon O_1^\prime
\end{equation}
as given correct the treatment of the axial gauge poles. We can now,
in principle, evaluate the effect of the $O_1^\prime$ term by looking
at the new effective quadratic form in (\ref{eq:presp6}). 
This turns out to be a more cumbersome procedure. We proceed 
along an alternate route as below.

Consider the effect of the net $\epsilon$-term in (\ref{eq:presp6}) on an
axial gauge Green's function:
\begin{equation}
\label{eq:presp7}
\langle O\rangle_A=\int{\cal D}\phi^\prime O[\phi^\prime]
e^{iS^A_{\rm eff}[\phi^\prime]}.
\end{equation}
This is given by the modification $S^A_{\rm eff}\rightarrow {S^A_{\rm eff}}\ ^\prime$ in 
(\ref{eq:presp7}),viz:
\begin{equation}
\label{eq:presp8}
\langle O\rangle_A=\int{\cal D}\phi^\prime O[\phi^\prime]
e^{iS^A_{\rm eff}+i\epsilon O_1^\prime}.
\end{equation}
We now proceed to relate
(\ref{eq:presp8}) to the corresponding Green's functions
in Lorentz gauges as done in \cite{BRSpap1}. We reexpress $\langle O\rangle_A$ as:
\begin{eqnarray}
\label{eq:presp9}
& & \langle O\rangle_A={1\over i}
{\delta\over {\delta N}}\int {\cal D}\phi^\prime e^{iS^A_{\rm eff}[\phi^\prime]
+i\epsilon O_1^\prime[\phi^\prime]+iN
O[\phi^\prime]}|_{N=0}\nonumber\\
& & \equiv{1\over i}{\delta\over{\delta N}}
W^A[N]|_{N=0}.
\end{eqnarray}
Now, we consider the quantity:
\begin{eqnarray}
\label{eq:presp10}
& & W^A[N]
\equiv\int {\cal D}\phi^\prime e^{iS^A_{\rm eff}[\phi^\prime]}
.\{e^{i\epsilon O_1^\prime[\phi^\prime]+iN O[\phi^\prime]}\}\nonumber\\
& & \equiv
\int {\cal D}\phi^\prime e^{iS^A_{\rm eff}[\phi^\prime]}f[\phi^\prime]
\end{eqnarray}
We now apply the procedure of \cite{BRSpap1} (following equation (34) of
that work) to the above expression
where $O[\phi^\prime]$ there is replaced by the
curly bracket above. Then, using identity (53) of \cite{BRSpap1}, we 
obtain that
\begin{equation}
\label{eq:presp11}
W^A[N]=\int{\cal D}\phi e^{iS^L_{\rm eff}[\phi]}\{f[\phi]+\sum_i\delta\phi_i
[\phi]{\delta f\over{\delta\phi_i}}\}.
\end{equation}
Now,
\begin{eqnarray}
\label{eq:presp12}
& & f[\phi]+\sum_i\delta\phi_i[\phi]{\delta f\over{\delta\phi_i}} 
=exp\biggl[i\biggl(\epsilon(O_1^\prime+\sum\delta\phi_i[\phi]
{\delta O_1^\prime\over{\delta\phi_i}})\nonumber\\
& &  +N(O+\sum_i\delta\phi_i[\phi]{\delta O\over{\delta\phi_i}})\biggr)\biggr].
\end{eqnarray}
In writing
(\ref{eq:presp12}), we have used the nilpotency of $\Theta^\prime$ contained
in each of $\delta\phi_i[\phi]$ [See equation (\ref{eq:result1})]. Hence,
\begin{equation}
\label{eq:presp13}
W^A[N]=\int{\cal D}\phi e^{iS^L_{\rm eff}[\phi]}
e^{i(\epsilon[O_1^\prime+\sum\delta\phi_i[\phi]
{\delta O_1^\prime\over{\delta\phi_i}}]
+N[O+\sum_i\delta\phi_i[\phi]{\delta O\over{\delta\phi_i}}])}.
\end{equation}
Now at $N=0$, the above must coincide with $W^L$ of (\ref{eq:presp2}). Hence,
\begin{equation}
\label{eq:presp14}
\epsilon\biggl[O_1^\prime
+\sum_i\delta\phi_i[\phi]{\delta O_1^\prime\over{\delta\phi_i}}
\biggr]=\epsilon O_1[\phi]=-i\int d^4x\epsilon(A^2/2-{\bar c}c)
\end{equation}
Now, following the transition from (53) to
(54) of \cite{BRSpap1}, we can make a transition in (\ref{eq:presp10}) above.
This amounts to substitution of $\delta\phi_i$
via (\ref{eq:result1}) and (\ref{eq:Thetaprimedef}). (Here we note that 
this is possible because the
${\cal O}(\epsilon)$ terms in (\ref{eq:presp10}) are independent
of $\kappa$) We then have
\begin{eqnarray}
\label{eq:presp15}
& & \langle O\rangle_A = \langle O\rangle_L\nonumber\\
& & +\int_0^1 d\kappa\int{\cal D}\phi e^{i[S^M_{\rm eff}[\phi,\kappa]
-i\epsilon\int d^4x(A^2/2-{\bar c}c)]}\sum_i\biggl(\delta_{1,i}[\phi]
+\kappa\delta_{2,i}[\phi]\biggr)\Theta^\prime[\phi]
{\delta O\over{\delta\phi_i}}.\nonumber\\
& & 
\end{eqnarray}
Thus, while the ${\cal O}(\epsilon)$ terms needed to be calculated
in the Green's function calculation for (\ref{eq:presp7}) 
(viz $\epsilon O_1^\prime$) are very complicated),when $\langle O\rangle_A$ 
is reexpressed as an integral over $\kappa$, the effect
of $\epsilon$ terms in this integral is simply to modify
$S_M[\phi,\kappa]\rightarrow S_M[\phi,\kappa]+i\epsilon
\int d^4x(-A^2/2+{\bar c}c)$. Thus, the form (\ref{eq:presp15}) facilitates the
evaluation of the effect of ${\cal O}(\epsilon)$ terms on
the axial gauge Green's
functions, as the modification there is
$\kappa$-independent.

We shall now use (\ref{eq:presp15})  to write down the expression for the axial
gauge propagator. So we consider:
\begin{equation}
\label{eq:AA}
O[\phi]\equiv A^\alpha_\mu(x)A^\beta_\nu(y).
\end{equation}
Then, with obvious notations
\begin{eqnarray}
\label{eq:exres}
& & iG^{A\ \alpha\beta}_{\mu\nu}(x-y)=iG^{L\ \alpha\beta}_{\mu\nu}(x-y)
+i\int_0^1 d\kappa\int{\cal D}\phi e^{iS_{\rm eff}^M[\phi,\kappa]-i
\epsilon\int(A^2/2-{\bar c}c)d^4x}\nonumber\\
& & \times
\biggl(({\rm D}_\mu c)^\alpha(x)A^\beta_\nu(y)
+A^\alpha_\mu(x)
({\rm D}_\nu c)^\beta(y)\biggr)
\int d^4z{\bar c}(z)(\partial\cdot A^\gamma-\eta\cdot A^\gamma)(z)
\nonumber\\
& & 
\end{eqnarray}  
The above is an exact result for  the relation between propagators
valid  to all orders.  For
obtaining the correct treatment for the $1/(\eta\cdot q)$-singularity,
we  are however interested in the tree propagator. For this, we collect the 
$O(g^0)$ terms on the right-hand side. Noting
\begin{eqnarray}
\label{eq:ghprop}
& & \int {\cal D}\phi c^\alpha(x){\bar c}^\beta(y)
e^{i[S^M_{\rm eff}-i\epsilon\int(A^2/2-{\bar c}c)d^4x]}
=i\delta^{\alpha\beta}\tilde G^{0 M}(x-y)\nonumber\\
& & =i\delta^{\alpha\beta}\int d^4q 
{e^{-iq\cdot(x-y)}\over{[(\kappa-1)q^2-i\kappa q\cdot\eta -i\epsilon]}}.
\end{eqnarray}
We can write, for  the tree propagator
 $G^{0}_{\mu\nu}$:
\begin{eqnarray}
\label{eq:exres1}
&  & G^{0 A\ \alpha\beta}_{\mu\nu}(x-y)=G^{0 L\ \alpha\beta}_{\mu\nu}(x-y)
\nonumber\\
& & +i\int_0^1 d\kappa\biggl[-i\partial_\mu^x
\tilde G^{0M}(x-y)(\partial_z^\sigma-\eta^\sigma)
\tilde G^{0M\ \alpha\beta}_{\sigma\nu}
+(\mu,x,\alpha)\leftrightarrow(\nu,y,\beta)\biggr].
\nonumber\\
& &
\end{eqnarray} 

In the next section, we shall use the result (\ref{eq:exres1}) to obtain
the correct $\epsilon$-dependent propagator in the axial gauges.

\section{Evaluation of the Axial Gauge Propagator}

In this section , we shall evaluate the $\epsilon$-dependent axial gauge
propagator using (\ref{eq:exres1}). We shall show
that over most of the real $\eta\cdot k$ axial, we recover the 
naive axial gauge propagator. However, we find a non-trivial
complex structure in a small region near $\eta\cdot k=0$. 
In this section, we shall content
ourselves with the algebraic study of $G^{0 A}_{\mu\nu}$.
In the next section, we shall study, in detail, the analytic structure
of $G^{0 A}_{\mu\nu}$ over the complex $\eta\cdot k$ plane.

We express (\ref{eq:exres1}), in momentum space as:
\begin{eqnarray}
\label{eq:exres2}
&  & G^{0 A}_{\mu\nu}(k)=G^{0 L}_{\mu\nu}(k)
\nonumber\\
& & +i\int_0^1 d\kappa\biggl[k_\mu^x
\tilde G^{0M}(k,\kappa)(-ik^\sigma-\eta^\sigma)
\tilde G^{0M}_{\sigma\nu}(k,\kappa)
+(\mu,k)\leftrightarrow(\nu,-k)\biggr].
\nonumber\\
& &
\end{eqnarray} 
Here 
\begin{equation}
\label{eq:eps5}
\tilde G^{0 M}(k,\kappa)
={1\over{(\kappa-1)k^2-i\kappa k\cdot\eta-i\epsilon}}
\end{equation}
and
\begin{equation}
\label{eq:exres3}
\tilde G^{0 M}_{\mu\rho}(k,\kappa)=Z^{-1}_{\mu\rho}
\end{equation}
with
\begin{equation}
\label{eq:exres4}
Z_{\mu\nu}\equiv
-(k^2+i\epsilon)\biggl[g_{\mu\rho}+\biggl({1\over\lambda}(1-\kappa)^2-1\biggr)
{k_\mu k_\rho\over{k^2+i\epsilon}}
-i{{\kappa(1-\kappa)}\over\lambda}{k_{[\mu}\eta_{\rho]}\over{k^2+i\epsilon}}
+{\kappa^2\over\lambda}{\eta_\mu\eta_\rho\over{k^2+i\epsilon}}\biggr],
\end{equation}
is the quadratic form in momentum space arising from the
$\epsilon$-dependent 
action $S^M_{\rm eff}-i\epsilon\int (A^2/2-{\bar c}c)$
in (\ref{eq:exres}). 
[It should be emphasized $\tilde G^{0M}_{\mu\nu}$ and
$\tilde G^{0M}$ are only
$intermediate$ objects occurring in calculations
and are $not$ the actual   ghost   and gauge
propagators  in the mixed gauges. For
example, while $\tilde G^{0M}_{\mu\nu}(k,\kappa=0)$ is equal to 
$\tilde G^{0 L}_{\mu\nu}$, 
$\tilde G^{0M}_{\mu\nu}(k,\kappa=1)\neq
G^{0A}_{\mu\nu}(k)$ because the latter has to be
evaluated with  the exact $\epsilon O_1^\prime$ terms in the
exponent and not $\epsilon(A^2/2-{\bar c}c)$ as occurring in
(\ref{eq:exres}). The $actual$ tree propagator in
mixed gauges would similarly be evaluated with
an appropriate term $\epsilon O_1[\phi,\kappa]$ 
and not from (\ref{eq:exres}); this is not required in our
evaluation.]
We express $\tilde G^{0M}_{\mu\rho}$ as:
\begin{eqnarray}
\label{eq:pres6}
& & \tilde G^{0 M}_{\mu\rho}(k)
=-{1\over{k^2+i\epsilon}}\Biggl[g_{\mu\rho}+\nonumber\\
& & {{\biggl(\biggl[[(1-\kappa)^2-\lambda]-{\eta^2\kappa^2\over{k^2+i\epsilon}}
\biggr]k_\mu k_\rho
-i\kappa(1-\kappa)k_{[\mu}\eta_{\rho]}
+{\kappa^2\eta\cdot k\over{k^2+i\epsilon}}k_{[\mu}\eta_{\rho]_+}
+i{\kappa^2\epsilon\over{k^2+i\epsilon}}\eta_\mu\eta_\rho\biggr)}
\over{\biggl(-{\kappa^2\over{(k^2+i\epsilon)}}
\biggl[(\eta\cdot k)^2-\eta^2k^2+(k^2+\eta^2)(k^2+i\epsilon)\biggr]
+2k^2\kappa
-i\epsilon\lambda-k^2\biggr)}}\Biggr].
\nonumber\\
& & 
\end{eqnarray}
\begin{eqnarray}
\label{eq:Result1}
& & \tilde G^{0 A}_{\mu\nu}-\tilde G^{0 L}_{\mu\nu}
={-i\over{(k^2+i\epsilon)^2(1-i\xi_1-i\xi_2)(1-i\xi_2+\xi_1^2+i\xi_2\xi_3)}}
\nonumber\\
& & \times\int_0^1 d\kappa{\Biggl[k_\mu k_\nu\biggl(\kappa
+\biggl[{i\lambda-\xi_1(1-\lambda)\over{\xi_1+i\xi_3}}\biggr]
\biggr)
(\xi_1+i\xi_3)+\eta_\mu k_\nu
\biggl(\kappa+\biggl[{1-i\xi_2(1-\lambda)\over{-1-i\xi_1+i\xi_2}}\biggr]\biggr)
(-1-i\xi_1+i\xi_2)\Biggr]
\over{(\kappa-a_1)(\kappa^2-2\gamma\kappa+\beta)}}\nonumber\\
& & +(k\rightarrow-k,\ \mu\leftrightarrow\nu) 
\end{eqnarray}
with
\begin{eqnarray}
\label{eq:defins}
& & \xi_1\equiv{\eta\cdot k\over{k^2+i\epsilon}};\nonumber\\
& & \xi_2\equiv{\epsilon\over{k^2+i\epsilon}};\nonumber\\
& & \xi_3\equiv{\eta^2\over{k^2+i\epsilon}};\nonumber\\
& & a_1\equiv{1\over{1-i\xi_1-i\xi_2}};\nonumber\\
& & \gamma\equiv{(1-i\xi_2)\over{1-i\xi_2+\xi_1^2+i\xi_2\xi_3}}\equiv{1-i\xi_2\over D};\nonumber\\
& & \beta\equiv{1+i\xi_2(\lambda-1)\over{1-i\xi_2+\xi_1^2+i\xi_2\xi_3}}
=\gamma +{i\xi_2\lambda\over D}
\end{eqnarray}
The quadratic in the denominator 
can be rewritten as $(\kappa-\kappa_1)(\kappa-\kappa_2)$ with
\begin{eqnarray}
\label{eq:kappa12def}
& & \kappa_{1,2}=\gamma\pm\sqrt{\gamma^2-\beta}
={{1-i\xi_2}\pm
\sqrt{(1-i\xi_2)^2
-[1+i\xi_2(\lambda-1)](1-i\xi_2+\xi_1^2+i\xi_2\xi_3)}\over D}\nonumber\\
& & \equiv{1-i\xi_2\pm\sqrt{Y}\over D}
\end{eqnarray}
We  note that of  the three zeros of the denominators, 
two are equal are $\epsilon=0$, since
\begin{eqnarray}
\label{eq:a1kappa1}
& & \kappa_1|_{\epsilon=0}
={1\over{1+\xi_1^2}}+\sqrt{\biggl(
{1\over{1+\xi_1^2}}\biggr)^2-{1\over{1+\xi_1^2}}}\nonumber\\
& & ={1+i\xi_1\over{1+\xi_1^2}}=a_1|_{\epsilon=0}.
\end{eqnarray}
We shall now state an important convention in defining the square roots in
(\ref{eq:kappa12def}). The square root $\sqrt{Y}$ has branch points 
at $\pm\sqrt{{-i\xi_2[(1-i\xi_2)\xi_3+\lambda(1-i\xi_2+i\xi_2\xi_3)]
\over{1-i\xi_2(1-\lambda)}}}$ and these lie a distance O$(\sqrt{\epsilon})$
away from the origin [For LCG, in the $k^2=0$ subspace, 
however, $\sqrt{Y}=i\sqrt{\lambda}\xi_1$
has no branch cut in $\xi_1$-plane]. We choose
the branch cut joining these. To obtain the value of 
$+\sqrt{Y}$ at any point $\xi^\prime$ 
not on the branch cut, we consider $\sqrt{Y}$
for $\xi_1=M\xi^\prime$ as $M\to+\infty$. 
Then we can ignore $\epsilon$ terms in this
case and $\sqrt{Y}=\sqrt{-\xi_1^2}$. This, we define to be 
$i\xi_1$. We then define $\sqrt{Y}$ for $\xi_1=\xi^\prime$ by requiring that
the phase of $\sqrt{Y}$ is a continuous function of $M$ for $1\leq M<\infty$.
>From this and from the fact that $Y\equiv Y(\xi_1^2)$, we learn that
$\sqrt{Y}(-\xi_1)=-\sqrt{Y}(\xi_1)$. Hence, $\kappa_2(-\xi_1)=\kappa_1(\xi_1)$.
We further note that this prescription defines
uniquely $\sqrt{Y}$ for  real $\eta\cdot k\neq 0$ since  the branch cut  cuts
the real $\eta\cdot k$ axis only at the
origin $\eta\cdot k=0$.

We further note that both the $k_\mu k_\nu$
and the $\eta_\mu k_\nu$ terms involve an integral of the same form
\begin{equation}
\label{eq:kappaint}
\int_0^1 d\kappa 
{(\kappa+\alpha)\over{(\kappa-a_1)(\kappa-\kappa_1)(\kappa-\kappa_2)}},
\end{equation}
the constant $\alpha$ being different for the $k_\mu k_\nu$ and 
$\eta_\mu k_\nu$ terms.
This can be evaluated and reorganized as:
\begin{eqnarray}
\label{eq:pres10}
& & {(a_1+\alpha)\over{(a_1-\kappa_1)(a_1-\kappa_2)}}
ln\biggl[{{1-a_1}\over{-a_1}}\biggr]\nonumber\\
& & +{(\kappa_1+\alpha)\over{(\kappa_1-a_1)(\kappa_1-\kappa_2)}}
ln\biggl[{{1-\kappa_1}\over{-\kappa_1}}\biggr]
+{(\kappa_2+\alpha)\over{(\kappa_2-a_1)
(\kappa_2-\kappa_1)}}ln\biggl[{{1-\kappa_2}\over{-\kappa_2}}\biggr]
\nonumber\\
& & \equiv {1\over{a_1-\kappa_2}}\biggl({(\alpha+a_1)\over{(a_1-\kappa_1)}}
ln\biggl[{{\kappa_1-\kappa_1a_1}\over{a_1-a_1\kappa_1}}\biggr]
-{(\kappa_2+\alpha)\over{(\kappa_1-\kappa_2)}}
ln\biggl[{{\kappa_2-\kappa_1\kappa_2}
\over{\kappa_1-\kappa_1\kappa_2}}\biggr]\biggr).
\end{eqnarray}
It is shown in Appendix A that the contribution of the second term
vanishes in the limit $\epsilon\to 0$.
Hence the propagator (\ref{eq:Result1}) is given in terms of the first term
in (\ref{eq:pres10}):
\begin{equation}
\label{eq:mainln}
{a_1+\alpha\over{(a_1-\kappa_2)(a_1-\kappa_1)}}
ln\biggl[{{\kappa_1-\kappa_1a_1}\over{a_1-a_1\kappa_1}}\biggr]
\end{equation}
substituted for (\ref{eq:kappaint}) in (\ref{eq:Result1}). Hence, we shall study the
structure of (\ref{eq:mainln}) in detail.
The singularity structure of (\ref{eq:mainln}) is dependent on the 
denominators and the logarithm. The equation
(\ref{eq:mainln}), in general reads:
\begin{equation}
\label{eq:lnP}
{(a_1+\alpha)D(1-i\xi_1-i\xi_2)^2
\over{i\xi_2(1-\lambda)P(\xi_1)}}ln\biggl[{-i(\xi_1+\xi_2)
\over{{-i\xi_2\lambda\over{1-i\xi_2(1-\lambda)}}
-\sqrt{[{1-i\xi_2\over{1-i\xi_2(1-\lambda)}}]^2-{D\over{1-i\xi_2(1-\lambda)}}}}}
\biggr]
\end{equation}
with
\begin{equation}
\label{eq:defP}
P(\xi_1)\equiv
\xi_1^2+2i\xi_1(1-i\xi_2)+{\lambda+i\xi_2(1-2\lambda)
+\xi_2^2(1-\lambda)+\xi_3\over{1-\lambda}}.
\end{equation}
The apparent complexity of (\ref{eq:lnP})
actually exists only in the small region of the $\eta\cdot k$ complex
plane near the  origin. We note  that for 
$|a_1-\kappa_1|<|a_1(1-\kappa_1)|$, the 
expression
(\ref{eq:mainln}) can be expressed as 
\begin{equation}
\label{eq:expln}
{1\over{a_1-\kappa_1}}
ln\biggl[{\kappa_1-a_1\kappa_1\over{a_1-a_1\kappa_1}}\biggr]
=-{1\over{a_1-a_1\kappa_1}}
+O(a_1-\kappa_1).
\end{equation}
The condition $|a_1-\kappa_1|<|a_1(1-\kappa_1)|$ implies
\begin{equation}
\label{eq:condition}
Im\biggl({\sqrt{-(\eta\cdot k)^2-i\epsilon\eta^2}\over
{\sqrt{{k^2\over{k^2+i\epsilon}}}(\eta\cdot k+\epsilon)}}\biggr)>{1\over 2}
\end{equation}
and this covers all of real $\eta\cdot k$ axis save the region
$(-\epsilon,0)$ for $\eta^2\neq0$ and $(-\epsilon,\epsilon)$ for LCG. Thus,
(\ref{eq:expln}) reads neglecting O$(\epsilon)$ terms
\begin{equation}
\label{eq:expln2}
-{1\over{a_1(1-\kappa_1)(a_1-\kappa_2)}}.
\end{equation}
For $|\eta\cdot k|>>\epsilon$, this is easily seen to be
\begin{equation}
\label{eq:smalln.k}
-{(1-i\xi_1)^2(1+\xi_1^2)\over{2\xi_1^2}}
\end{equation}
and leads
 to the usual behavior of the axial propagator
when substituted into
(\ref{eq:Result1}), which then reads (See Appendix C):
\begin{equation}
\label{eq:eps=0}
\tilde G^{0 A}_{\mu\nu}-\tilde G^{0 L}_{\mu\nu}= 
-{1\over k^2}k_\mu k_\nu
\biggl({(\lambda k^2+\eta^2)\over{(\eta\cdot k)^2}}+{(1-\lambda)\over k^2}
\biggr) + {k_{[\mu}\eta_{\nu]_+}\over{k^2\eta\cdot k}}.
\end{equation}

We finally summarize our results. We find:
\begin{equation}
\label{eq:freslt1}
\tilde G^{0 A}_{\mu\nu}=\tilde G^{0 L}_{\mu\nu}\
+\biggl[\biggl(k_\mu k_\nu\Sigma_1+\eta_\mu k_\nu\Sigma_2\biggr)ln\Sigma_3
+(k\to -k;\mu\leftrightarrow\nu)\biggr]
\end{equation}
where
\begin{eqnarray}
\label{eq:freslt2}
& & \Sigma_1\equiv {-(k^2-i\eta\cdot k)\biggl({\eta\cdot k+i\eta^2
\over{k^2-i\eta\cdot k}}+i\lambda
-{(1-\lambda)\eta\cdot k\over{k^2+i\epsilon}}\biggr)
\over{\epsilon\Sigma}}\nonumber\\
& & \Sigma_2\equiv {-(k^2-i\eta\cdot k)\biggl(
-\biggl[{k^2+i\eta\cdot k\over{k^2-i\eta\cdot k}}
\biggr]+1-{i\epsilon(1-\lambda)\over{k^2+i\epsilon}}\biggr)
\over{\epsilon\Sigma}}\nonumber\\
\nonumber\\
& & \Sigma_3\equiv{-i(\eta\cdot k+\epsilon)(k^2+i\epsilon\lambda)
\over{(k^2+i\epsilon)\biggl(-i\epsilon\lambda-\sqrt{k^4-(k^2+i\epsilon\lambda)
\biggl[k^2+{(\eta\cdot k)^2+i\epsilon\eta^2\over{k^2+i\epsilon}}\biggr]}\biggr)}},\nonumber\\
& & {\rm and}\nonumber\\
& & \Sigma\equiv\biggl[(1-\lambda)[(\eta\cdot k)^2+2ik^2\eta\cdot k]
+i\epsilon k^2(1-2\lambda)+\lambda(k^2+i\epsilon)^2+\eta^2(k^2+i\epsilon)\biggr].
\nonumber\\
& & 
\end{eqnarray} 

\section{Singularity  Structure of the Propagator}

In  this section, we shall
study the singularity structure of the
propagator both on the real  $\eta\cdot k$ axis
as well as the $\eta\cdot k$ complex plane in general.  
The singularity structure on  the  real axis is important from
the point of vie w of  the well-defined  nature  of the
propagator for real $k_\mu$ while the singularity 
structure in the complex $\eta\cdot k$ plane is  relevant the
question of Wick rotation.

As shown in Section 4,  the quantity (\ref{eq:lnP}) is relevant
to both the propagator terms of  $k_\mu k_\nu$ and
$(k_\mu\eta_\nu+k_\nu\eta_\mu)$ kind. We shall first analyze its
structure. The singularities of (\ref{eq:lnP}) arise from
those of $P(\xi_1)$ and from those of $ln$ term. We shall
first analyze the singularities of  $P(\xi_1)$.

$P(\xi_1)$ is  a   quadratic polynomial in 
$\xi_1={\eta\cdot k\over{k^2+i\epsilon}}$.
It has two zeros; they are:
\begin{equation}
\label{eq:zerosP1}
 \xi_1=-i(1-i\xi_2)\pm i\sqrt{{1-i\xi_2+\xi_3\over{1-\lambda}}},
\end{equation}
i.e. at
\begin{equation}
\label{eq:zerosP2}
\eta\cdot k=-ik^2\pm i\sqrt{{(k^2+\eta^2)(k^2+i\epsilon)\over{1-\lambda}}}.
\end{equation}
We note that the above roots vanish only when $k^2$ satisfies:
\begin{equation}
\label{eq:zerosP3}
\lambda k^4+k^2(\eta^2+i\epsilon)+i\epsilon\eta^2=0.
\end{equation}
The above equation for $k^2=0$ has physical (i.e. real $k^2$)
root(s) only in $\eta^2=0$ (except for $\lambda=1$). For $\eta^2=0$,
it is easily seen that both the roots in (\ref{eq:zerosP2})
vanish at $k^2=0$. We thus note that in the 
light cone gauge ($\eta^2=0$), the point $k^2=0$ needs
to be treated with care. We shall first discuss the case $\eta^2\neq0$.

\underline{Case I: $\eta^2>0$}

(i) $k^2>0$: Hence, $(k^2+\eta^2)k^2>k^4>0$. We may set $\epsilon=0$. The roots
are:
\begin{equation}
\label{eq:zerosP4}
\eta\cdot k=-ik^2\biggl[1\mp\sqrt{{1\over{1-\lambda}}\biggl(1+{\eta^2\over k^2}\biggr)}\biggr]
\end{equation}
and lie on imaginary axis on wither side of the real axis at a finite
distance even as $\epsilon\to 0$.

(ii) $k^2=0$: The roots are at
\begin{equation}
\label{eq:zerosP5}
\eta\cdot k=\pm i\sqrt{{i\eta^2\epsilon\over{1-\lambda}}}=\pm e^{{3i\pi\over4}}
\sqrt{{\eta^2\epsilon\over{1-\lambda}}}.
\end{equation}
The roots  again lie on either side of the real axis at an
infinitesimal distance on the line $\theta={3\pi\over 4}$.

(iii) $0>k^2>-\eta^2$: Here $(k^2+\eta^2)k^2<0$. We can therefore
write the roots as:
\begin{equation}
\label{eq:zerosP6}
\eta\cdot k=-ik^2\pm\sqrt{{|(k^2+\eta^2)k^2|-i\epsilon|k^2+\eta^2|
\over{1-\lambda}}}.
\end{equation}
Again, we may set $\epsilon=0$ here.
Then, the roots
\begin{equation}
\label{eq:zerosP7}
\eta\cdot k=-ik^2\pm\sqrt{{|(k^2+\eta^2)k^2|\over{1-\lambda}}}
\end{equation}
are on the same side of the real axis, i.e. in the the upper half plane (UHP)
for $k^2<0$.

(iv) $k^2<-\eta^2$:
Here $0<k^2(k^2+\eta^2)<k^4$. Then (setting $\epsilon=0$), the roots
are at:
\begin{equation}
\label{eq:zerosP8}
\eta\cdot k=i|k^2|\biggl[1\pm\sqrt{{1\over{\lambda-1}}\biggl(1
-{\eta^2\over{|k^2|}}}\biggr)\biggr],
\end{equation}
and both lie in the UHP.

\underline{Case II: $\eta^2<0$}

We express (\ref {eq:zerosP2}) as
\begin{equation}
\label{eq:zerosP9}
\eta\cdot k=i(-k^2)\pm i\sqrt{{(-k^2-\eta^2)(-k^2-i\epsilon)\over{1-\lambda}}}.
\end{equation}
We note that the above expression is analogous to the RHS 
 of (\ref{eq:zerosP2}) with
$-\eta^2>0$ and $k^2\to -k^2,\ \epsilon\to-\epsilon$. Hence,
a discussion parallel to that given above applies. We find:

(i) $k^2<0$: The poles are at 
\begin{equation}
\label{eq:zerosP10}
\eta\cdot k=i|k^2|\biggl[1\pm\sqrt{{1\over{1-\lambda}}\biggl(1+{|\eta^2|
\over |k^2|}\biggr)}\biggr]
\end{equation}
and lie on imaginary axis on wither side of the real axis at a finite
distance even as $\epsilon\to 0$.

(ii) $k^2=0$
\begin{equation}
\label{eq:zerosP11}
\eta\cdot k=\pm i\sqrt{{-i|\eta^2|\epsilon\over{1-\lambda}}}=\pm e^{{i\pi\over4}}
\sqrt{{|\eta^2|\epsilon\over{1-\lambda}}}.
\end{equation}
The roots  again lie on either side of the real axis at an
infinitesimal distance on the line $\theta={\pi\over 4}$.

(iii) $0<k^2<-\eta^2$: Here $(k^2+\eta^2)k^2<0$. We can therefore
write the roots as:
\begin{equation}
\label{eq:zerosP12}
\eta\cdot k=ik^2\pm\sqrt{{|(k^2+\eta^2)k^2|\over{1-\lambda}}}
\end{equation}
and  both lie in the UHP.

(iv) $k^2<-\eta^2$:
Here $0<k^2(k^2+\eta^2)<k^4$. Then (setting $\epsilon=0$), the roots
are at:
\begin{equation}
\label{eq:zerosP13}
\eta\cdot k=-ik^2\biggl[1\mp \sqrt{{1\over{\lambda-1}}\biggl(1
-{|\eta^2|\over{|k^2|}}}\biggr)\biggr],
\end{equation}
and  both lie in the LHP.

Finally, we consider the important case of the light
cone gauge (LCG). It is this case, where
 most of the difficulties associated with the 
prescriptions obtained by others are located.

\underline{Case III: $\eta^2=0$}
 We note that the roots of (\ref{eq:zerosP2})
now read:
\begin{equation}
\label{eq:zerosP14}
\eta\cdot k=-ik^2\pm i\sqrt{{k^4+i\epsilon k^2\over{1-\lambda}}}.
\end{equation}
For $k\neq0$, we can write the roots as:
\begin{equation}
\label{eq:zerosP15}
\eta\cdot k=-ik^2\biggl[1\mp\sqrt{{1\over{1-\lambda}}\biggl(1+i{\epsilon
\over k^2}\biggr)}\biggr].
\end{equation}
Without loss of generality, we may assume that
$\epsilon<|k^2|$ and expand the square root. The we find:
\begin{equation}
\label{eq:zerosP16}
\eta\cdot k=-ik^2\biggl[1\mp{1\over{\sqrt{1-\lambda}}}\mp
{1\over{8\sqrt{1-\lambda}}}{\epsilon^2\over k^4}\biggr]\mp
{\epsilon\over{2\sqrt{1-\lambda}}}.
\end{equation}
We note that for $0<\lambda<1$, the roots lie on either side of the real axis.
In particular for small positive $\lambda$, we may set $\epsilon=0$ and find:
\begin{equation}
\label{eq:zerosP17}
\eta\cdot k=-2ik^2,\ {i\lambda k^2\over 2}.
\end{equation}
On the other hand, we may set $\lambda=0$ in
the beginning. Then we find it necessary to
take into account the $O(\epsilon^2)$ term in (\ref{eq:zerosP15}).
We then find:
\begin{equation}
\label{eq:zerosP18}
\eta\cdot k=-2ik^2,\ {i\epsilon\over{8k^2}}.
\end{equation}
Thus, in either case, the roots are at:
(i) $\eta\cdot k=-2ik^2$, and
(ii) on imaginary axis, in UHP for $k^2>0$
and  LHP for $k^2<0$.
We note that this discussion clearly fails for $k^2=0$
in the LCG. We then necessarily have 
to obtain the treatment for the LCG by a limiting procedure $\eta^2\to0$
for this subspace of momenta. From (\ref{eq:zerosP2}),
for $k^2=0$, the roots $\eta\cdot k$ are:
\begin{equation}
\label{eq:zerosP19}
\eta\cdot k=\pm\sqrt{{-i\epsilon\eta^2\over{1-\lambda}}}.
\end{equation}
For $\eta\neq0$, (irrespective of where $\eta^2>0$ or $<0$)
these lie on the opposite sides of the
real axis at an infinitesimal distance away.

\section{Effective Treatment for the Axial Propagator}

We obtained an exact $\epsilon$-dependent expression for the axial gauge propagator
from its connection to the Lorentz gauge Green's functions.  As remarked
earlier, the propagator is effectively the same
as the usual one except in  a small region near $\eta\cdot k=0$.
It is in this region that the many 
treatments of the propagator have been suggested, differ. The expression
we have obtained ab initio, however, has complicated structure in this
region. We wish to show that it can be replaced by a much simpler
expression which yields
the same (coordinate space) propagator and will facilitate the 
actual axial gauge calculations rather than using the expressions of
(\ref{eq:freslt1}) and (\ref{eq:freslt2}).

We consider the tree axial gauge propagator in coordinate space:
\begin{eqnarray}
\label{eq:simpeq1}
& & \Delta_{\mu\nu}(x-y)=\int{d^4k\over{(2\pi)^4}}
e^{-ik\cdot(x-y)}G^{0A}_{\mu\nu}(k)
\nonumber\\
& & =\int{d^3k\over{(2\pi)^3}}e^{i\vec k\cdot(\vec x-\vec y)}
\int{dk^0\over{2\pi}}e^{-ik^0(x^0-y^0)}G^{0A}_{\mu\nu}(k)
\end{eqnarray}
We introduce the variable:
\begin{equation}
\label{eq:simpeq2}
\zeta\equiv\eta^0k^0-\vec\eta\cdot\vec  k.
\end{equation}
We shall deal with the case $\eta^0\neq0$. (This
is always possible by a proper choice of a Lorentz frame.) We further assume
for simplicity of treatment , $\eta^0=1$. (This can always
by arranged by rescaling $\eta_\mu$ if necessary.) Thus,
we use:
$\zeta=k^0-\vec\eta\cdot\vec k$ and $\vec k$ as
integration variables. We express
\begin{equation}
\label{eq:simpeq3}
\Delta_{\mu\nu}(x-y)=\int{d^3k\over{(2\pi)^3}}
e^{i\vec k\cdot(\vec x-\vec y)-i\vec\eta\cdot\vec k(x^0-y^0)}
\int{d\zeta\over{(2\pi)}}e^{-i\zeta(x^0-y^0)}G^{0A}_{\mu\nu}(\zeta,\vec k).
\end{equation}
We shall focus our attention on those terms in 
$G^{0A}_{\mu\nu}(k)=G^{0L}_{\mu\nu}
+[G^{0A}_{\mu\nu}(k)-G^{0L}_{\mu\nu}(k)]$ which have nontrivial
structure near $\zeta=0$. We shall in effect show that
\begin{equation}
\label{eq:simpeq4}
\int_{-\infty}^\infty
{d\zeta\over{(2\pi)}}
e^{-i\zeta(x^0-y^0)}G^{0A}_{\mu\nu}(\zeta,\vec k)
\end{equation}
can, in the limit $\epsilon\to0$, be effectively replaced by two terms: (i) one
of which is
\begin{equation}
\label{eq:simpeq5}
\int_C{d\zeta\over{(2\pi)}}
e^{-i\zeta(x^0-y^0)}G^{0A}_{\mu\nu,{\rm eff}}(\zeta,\vec k)
\end{equation}
where $G^{0A}_{\mu\nu,{\rm eff}}$ has a simple
structure near $\zeta=0$ and the contour
$C$ is a contour suitably distorted near $\zeta=0$ as will
be specified soon, (ii) and a contribution having
a relatively simple form arising from the region near $\zeta=0$.

To achieve this, we replace,
\begin{equation}
\label{eq:simpeq6}
\int_{-\infty}^\infty=\int_Cd\zeta+\oint_{C_1}d\zeta
\end{equation}
where $C$ runs from $(-\infty,-\alpha\sqrt{\epsilon})$ and
$(\alpha\sqrt{\epsilon},\infty)$ and is completed by adding a
semicircle of radius $\alpha\sqrt{\epsilon}$ in the LHP. Contour $C_1$
on the other hand is a closed contour to compensate[C for the left hand 
side (See Fig 1). Here $\alpha$ is a (large enough) arbitrary positive
number. We then show that (i) on $C$, $G^{0A}_{\mu\nu}(\zeta,\vec k)$, the 
$\epsilon$-dependence can, in  fact, be ignored and be replaced by the naive
axial propagator (with $\zeta$ complex over the semicircle);
(ii) The contour $C_1$ can be shrunk so that the contribution over $C_1$ can be
replaced by that around the branch cut (from $-i\sqrt{i\epsilon\eta^2}$ to
$i\sqrt{i\epsilon\eta^2}$) which will then be evaluated (See Fig. 2). 
This contribution,
in the limit $\epsilon\to0$, can be replaced by a simple expression as shown
later.

[Here, we clarify the location of the contour on the left hand side
of (\ref{eq:simpeq6}). We recall that in (\ref{eq:freslt1}), the $ln$
factor:
\begin{equation}
\label{eq:lnfac1}
ln\biggl({\zeta+\epsilon\over{\sqrt{\zeta^2+i\epsilon\eta^2}}}\biggr).
\end{equation}
is in principle, multi-valued.
However, this factor, has arisen out of 
the expression (\ref{eq:mainln}) whose value is unambiguous in 
in the original complex integral[C (\ref{eq:kappaint}). In particular,
we have already defined $\sqrt{\zeta^2+i\epsilon\eta^2}$ earlier for
$|\zeta|>>\epsilon$, the factor boils down to
\begin{equation}
\label{eq:lnfac2}
ln\biggl({\zeta\over{i\zeta}}\biggr)=-i{\pi\over2}.
\end{equation}
In the region $|\zeta|\sim\epsilon$, we
must define the phase of the logarithm  recalling that it has 
arisen from the unambiguous integrals
\begin{eqnarray}
\label{eq:lnfac3}
& & \int_0^1{d\kappa\over{\kappa-a_1}}=ln\biggl({1-a_1\over{-a_1}}\biggr)
\\
& & \int_0^1{d\kappa\over{\kappa-\kappa_1}}=ln\biggl({1-\kappa_1\over{-\kappa_1}}\biggr).
\end{eqnarray}
This defines the way the contour on the left hand side of 
(\ref{eq:simpeq6}) should be drawn near $\zeta=0,\ -\epsilon$.]

In this work, we shall  only deal with the case $\eta^2\neq0$ and consider the
propagator for $k^2\neq0$. We may then choose $\epsilon<<|k^2|$, in which
case we may replace $k^2/(k^2+i\epsilon)$ by 1 wherever possible. We shall also set
$\lambda=0$for simplicity. [In the case of LCG, we may need to keep 
$\lambda\neq0$ till the end; we shall not deal with this here however.]

On the contour $C$, we have $|\zeta|>>\sqrt{\epsilon\eta^2}$ for sufficiently
large $\alpha$. Here we can employ the treatment in Appendices
B and C to conclude that the propagator over $C$ can be replaced
by the naive axial propagator (with complex $\zeta$):
\begin{equation}
\label{eq:naiveaxprop}
G^{0A}_{\mu\nu}(k,\epsilon=0)=-{1\over k^2}\biggl(g_{\mu\nu}
+{{(\lambda k^2+\eta^2)}\over{(\eta\cdot k)^2}}k_\mu k_\nu
-{k_{[\mu}\eta_{\nu]_+}\over{\eta\cdot k}}\biggr).
\end{equation}

We recall from the discussion of Section  4, that for $k^2\neq0,\eta^2\neq0$,
the singularities of $P(\xi_1)$ are at a finite distance away from $\zeta=0$. 
Hence $C_1$ does not enclose these. 
We however, have the branch cut defined earlier (see discussion following
(\ref{eq:a1kappa1})) and ones arising from the presence of the
logarithm. We shrink the contour $C_1$ so that  it in effect goes around
the branch cut in the LHP, and receives a contribution proportional to the
discontinuity in $ln(\sqrt{\zeta^2+i\epsilon\eta^2})$ across the branch
cut.

We summarize the procedure followed for evaluating the effective replacement.
We express 
\begin{eqnarray}
\label{eq:simpeq7}
& & 
\int{d^3k\over{(2\pi)^3}}e^{i\vec k\cdot(\vec x-\vec y)}
\oint_{C_1}
{dk^0\over{(2\pi)}}e^{-ik^0(x^0-y^0)}\biggl[k_\mu k_\nu A_1(\zeta,\vec k)
+\eta_\mu k_\nu A_2(\zeta,\vec k)+\eta_\nu k_\mu 
A_3(\zeta,\vec k)\biggr]\nonumber\\
& & =-\partial^x_\mu\partial^x_\nu
\int{d^3k\over{(2\pi)^3}}e^{i\vec k\cdot(\vec x-\vec y)
-i\vec\eta\cdot\vec k(x^0-y^0)}\oint_{C_1}{d\zeta\over{(2\pi)}}
e^{-i\zeta(x^0-y^0)}A_1(\zeta,\vec k)\nonumber\\
& & +i\eta_\mu\partial^x_\nu
\int{d^3k\over{(2\pi)^3}}e^{i\vec k\cdot(\vec x-\vec y)
-i\vec\eta\cdot\vec k(x^0-y^0)}\oint_{C_1}{d\zeta\over{(2\pi)}}
e^{-i\zeta(x^0-y^0)}A_2(\zeta,\vec k)\nonumber\\
& & +i\eta_\nu\partial^x_\mu
\int{d^3k\over{(2\pi)^3}}
e^{i\vec k\cdot(\vec x-\vec y)-i\vec\eta\cdot\vec k(x^0-y^0)}
\oint_{C_1}{d\zeta\over{(2\pi)}}
e^{-i\zeta(x^0-y^0)}A_3(\zeta,\vec k)
\end{eqnarray}
We then evaluate 
\begin{equation}
\label{eq:simpeq8}
\oint_{C_1} 
{d\zeta\over{(2\pi)}}e^{-i\zeta(x^0-y^0)}A_i(\zeta,\vec k)
\end{equation}
by replacing $C_1$ by a contour that goes around the branch cut as
mentioned earlier. The contribution comes only from the discontinuity of $A_i$
across the branch cut, and is equal to
 \begin{equation}
\label{eq:simpeq9}
\int_0^{-i\sqrt{i\epsilon\eta^2}}
{d\zeta\over{(2\pi)}}e^{-i\zeta(x^0-y^0)}{\rm Disc}A_i(\zeta,\vec k)
\end{equation}
We expand Disc$A_i(\zeta,\vec k)$ around $\zeta=0$:
\begin{equation}
\label{eq:simpeq10}
{\rm Disc} A_i(\zeta,\vec k)={1\over\epsilon}\sum_n a_{i(n)}\zeta^n
\end{equation}
where we have  explicitly shown the ${1\over\epsilon}$ dependence 
of the discontinuity  
We have:
\begin{eqnarray}
\label{eq:simpeq11}
& & 
{1\over\epsilon}\int_0^{-i\sqrt{i\epsilon\eta^2}}{d\zeta\over{(2\pi)}}
e^{-i\zeta(x^0-y^0)}\sum_{n=0}^\infty a_{i(n)}\zeta^n
\nonumber\\
& & = 
{1\over\epsilon}\int_0^{-i\sqrt{i\epsilon\eta^2}}{d\zeta\over{(2\pi)}}
\biggl(a_{i(0)}[1-i\zeta(x^0-y^0)]+a_{i(1)}\zeta\biggr)+O(\sqrt{\epsilon})
\nonumber\\
& & = -a_{i(0)}{1\over{2\pi}}i\sqrt{{i\eta^2\over\epsilon}}
e^{-\sqrt{i\epsilon\eta^2}(x^0-y^0)\over 2}
-{i\eta^2\over{2(2\pi)}}a_{i(1)}+O(\sqrt{\epsilon})
\nonumber\\
& & 
={e^{-\sqrt{i\epsilon\eta^2}(x^0-y^0)\over 2}\over{2\pi}}
\biggl[-i\sqrt{{i\eta^2\over\epsilon}}a_{i(0)}
-{i\eta^2\over 2}a_{i(1)}\biggr]+O(\sqrt{\epsilon})\nonumber\\
& & =\int_{-\infty}^\infty{d k^0\over{2\pi}}e^{-ik^0(x^0-y^0)}
\delta\biggl(k^0-{1\over2}\sqrt{{\epsilon\eta^2\over i}}\biggr)
\biggl[-i\sqrt{{i\eta^2\over\epsilon}}a_{i(0)}
-{i\eta^2\over 2}a_{i(1)}\biggr]+O(\sqrt{\epsilon})\nonumber\\
\end{eqnarray}
We can thus in effect replace the contribution in (\ref{eq:simpeq7}) by:
\begin{eqnarray}
\label{eq:simpeq12} 
& & \int{d^4k\over{(2\pi)^4}}e^{-ik\cdot(x-y)}\biggl\{\delta\biggl(k^0
-{1\over2}\sqrt{{\epsilon\eta^2\over i}}-\vec\eta\cdot\vec k\biggr)
\biggl(k_\mu k_\nu
\biggl[-i\sqrt{{i\eta^2\over\epsilon}}a_{1(0)}
-{i\eta^2\over 2}a_{1(1)}\biggr]\nonumber\\
& & +\eta_\mu k_\nu\biggl[-i\sqrt{{i\eta^2\over\epsilon}}a_{2(0)}
-{i\eta^2\over 2}a_{2(1)}\biggr]\nonumber\\
& & +\eta_\nu k_\mu\biggl[-i\sqrt{{i\eta^2\over\epsilon}}a_{3(0)}
-{i\eta^2\over 2}a_{3(1)}\biggr]\biggr)\biggr\};
\end{eqnarray}
the curly bracket above gives the effective addition to the propagator.

We
quote the values of $a_{i(n)},\ n=0,1;\ i=1,2,3$ for completeness:
\begin{eqnarray}
\label{eq:simpeq13}
& & a_{1(0)}=-{2i\eta^2\over{\cal K}_1}ln(-1)={2\pi\eta^2\over{\cal K}_1},
\nonumber\\
& & a_{1(1)}={i\eta^2[{\cal K}_2+k\rightarrow-k]\over{\cal K}_1^2}ln(-1)
={-4i\pi\eta^2\biggl[(\vec\eta\cdot\vec k)^2-\vec k^2\biggr]\over{\cal K}_1^2};
\nonumber\\
& & a_{2(0)}=0,\nonumber\\
& & a_{2(1)}={2i\over{\cal K}_1}ln(-1)=-{2\pi\over{\cal K}_1};
\nonumber\\
& & a_{3(0)}=0,\nonumber\\
& & a_{3(1)}=-{2i\over{\cal K}_1}ln(-1)={2\pi\over{\cal K}_1},
\end{eqnarray}
where 
\begin{eqnarray}
\label{eq:simpeq14}
& & {\cal K}_1\equiv\biggl((\vec\eta\cdot\vec k)^2-\vec k^2\biggr)
(\eta^2+i\epsilon);
\nonumber\\
& & {\cal K}_2\equiv 2i\biggl((\vec\eta\cdot\vec k)^2-\vec k^2\biggr)
+2\vec\eta\cdot\vec k(\eta^2+i\epsilon).
\end{eqnarray}

In the treatment we have given in this section, we have found it necessary to
keep $\eta^2$ nonzero in the intermediate stages. 
We can, however, define the LCG 
as a limit of the final result as $\eta^2\rightarrow 0$.
The procedure, here, is very
analogous to the one we had to adopt in Appendix A, where we found it
necessary to keep $\eta^2$ nonzero in the intermediate stages of calculations.
We need, however, take this limit keeping $\epsilon$ nonzero.
We find, by taking the limit 
$\eta^2\rightarrow0$ in (\ref{eq:simpeq12})-(\ref{eq:simpeq14}),
that these extra terms
vanish for LCG. This result can also be seen ab intio by looking at the contour 
integral over $C_1$. 
Here we note that the width of the branch cut shrinks as $\eta^2\rightarrow0$.
This together with the fact that [for $\epsilon$ nonzero] the discontinuity
is finite for $\eta^2=0$, leads to the above result.

\section{Conclusion}

In this work, we have dealt with the question of the treatment of 
$1/(\eta\cdot k)^p$-type singularities in
the axial/LCG gauges in an ab-initio manner.
We have used the known treatment in the Lorentz gauges and connection
between Lorentz and axial gauges
to achieve this. We have used the results established earlier on
the FFBRS transformation that connect Green's functions in these two
gauges. We evaluated  an $\epsilon$-dependent propagator
in axial gauges via this procedure. We have suggested
that this should give the  correct
way to deal with the $1/(\eta\cdot k)^p$-type 
singularities in the axial propagator.

We find that this propagator, not surprisingly,
coincides with the usual axial propagator except in a 
small region around $\eta\cdot k=0$, The propagator does not 
show spurious poles 
at $\eta\cdot k=0$. It is however complicated 
but {\it predetermined in form} in a small region near $\eta\cdot k=0$.
We have shown however that there is a way of effectively replacing
this complicated structure of the propagator by a much simpler expression.
We believe, this is a first ab-initio treatment of axial gauge poles in
the path-integral formalism.

As mentioned in the introducton and elaborated in [15], the prescription
obtained here also preserves the value of the Wilson loop.

\appendix
\section{}
\setcounter{equation}{0}
\seceqaa

In this appendix, we shall give the
complete treatment for the second term in (\ref{eq:pres10}). In particular, 
we shall  show that the contribution from this  
term vanishes as $\epsilon\to 0$.

We wish to consider the contribution of 
\begin{equation}
\label{eq:secterm}
{(\kappa_2+\alpha)\over{a_1-\kappa_1)}}
{1\over{(\kappa_1-\kappa_2)}}
ln\biggl[{{\kappa_2-\kappa_1\kappa_2}
\over{\kappa_1-\kappa_1\kappa_2}}\biggr]\biggr)
\end{equation}
to the integral
(\ref{eq:kappaint}) occurring in the propagator expression
(\ref{eq:Result1}). We, first, note that on account of  
$\kappa_2(-\xi_1)=\kappa_1(\xi_1)$, we have that under
$\xi_1\to-\xi_1$,
\begin{equation}
\label{eq:secterm2}
{1\over{(\kappa_1-\kappa_2)}}
ln\biggl[{{\kappa_2-\kappa_1\kappa_2}
\over{\kappa_1-\kappa_1\kappa_2}}\biggr]
\end{equation}
remains invariant.

Now, the contribution (\ref{eq:secterm}) to the $k_\mu k_\nu$ terms reads:
\begin{eqnarray}
\label{eq:secterm3}
& & k_\mu k_\nu{(\xi_1+i\xi_3)
\over{(k^2+i\epsilon)^2(1-i\xi_1-i\xi_2)(1-i\xi_2+\xi_1^2+i\xi_2\xi_3)}}
{\biggl[{i\lambda-\xi_1(1-\lambda)\over{\xi_1+i\xi_3}}
\biggr]\over{\biggl[{1\over{1-i\xi_1-i\xi_2}}-\kappa_2\biggr]}}\nonumber\\
& & \times{1\over{(\kappa_1-\kappa_2)}}
ln\biggl[{\kappa_2-\kappa_1\kappa_2\over{\kappa_1-\kappa_1\kappa_2}}\biggr]
+(\xi_1\rightarrow-\xi_1;\ \mu\leftrightarrow\nu)\nonumber\\
& & = -{k_\mu k_\nu
\over{(k^2+i\epsilon)^2(1-i\xi_2+\xi_1^2+i\xi_2\xi_3)}}
{1\over{(\kappa_1-\kappa_2)}}
ln\biggl[{{\kappa_2-\kappa_1\kappa_2}\over{\kappa_1-\kappa_1\kappa_2}}\biggr]
\nonumber\\
& & \times\biggl[
{(\xi_1+i\xi_3)\kappa_2+i\lambda-\xi_1(1-\lambda)
\over{\kappa_2(1-i\xi_1-i\xi_2)-1}}
+{(-\xi_1+i\xi_3)\kappa_2+i\lambda+\xi_1(1-\lambda)
\over{\kappa_1
(1+i\xi_1-i\xi_2)-1}}\biggr].\nonumber\\
& & 
\end{eqnarray}
The last square bracket in  (\ref{eq:secterm3}) can be simplified as:
\begin{equation}
\label{eq:secterm4}
{[(\xi_1+i\xi_3)\kappa_2+i\lambda-\xi_1(1-\lambda)]
[\kappa_1(1+i\xi_1-i\xi_2)-1]+(\xi\rightarrow-\xi_1)
\over{ 
[\kappa_1(1+i\xi_1-i\xi_2)-1][\kappa_2(1-i\xi_1-i\xi_2)-1]}}.
\end{equation}
We note  that the denominator is even under $\xi_1\rightarrow-\xi_1$.
The numerator is now evaluated keeping in mind that $\kappa_1\kappa_2$
and $\kappa_1+\kappa_2$ are even under $\xi_1\rightarrow-\xi_1$
and $(\kappa_1-\kappa_2)$ is odd. The net contribution of 
(\ref{eq:secterm4}) to the $k_\mu k_\nu$ terms in the
propagator reads:
\begin{equation}
\label{eq:secterm5}
-{k_\mu k_\nu\over{(k^2+i\epsilon)^2}} f(\eta\cdot k,k^2,\eta^2,\lambda)
\end{equation}
with
\begin{eqnarray}
\label{eq:deff}
& & 
f\equiv{1\over{\sqrt{\gamma^2-\beta}}}
ln\biggl[{{\kappa_2-\kappa_1\kappa_2}\over{\kappa_1-\kappa_1\kappa_2}}\biggr]
{1\over{1-i\xi_2+\xi_1^2+i\xi_2\xi_3}}\nonumber\\
& & \times{i\xi_1^2(\beta-\gamma+\lambda\gamma)+i\xi_3[(\beta-\gamma)-i\xi_2\beta]-i\lambda[1-\gamma(1-i\xi_2)]
+\sqrt{\gamma^2-\beta}i\xi_1\xi_2(1-\lambda)
\over{[\kappa_1(1+i\xi_1-i\xi_2)-1][\kappa_2(1-i\xi_1-i\xi_2)-1]}}.\nonumber\\
& & 
\end{eqnarray}

We note:

(i) for $k^2\neq0$, $\kappa_2=1/(1+i\xi_2)=\kappa^\ast$ at
$\epsilon=0$;

(ii) the denominator is well defined at $\epsilon=0$, for $\eta\cdot k\neq 0$,
$k^2\neq0$;

(iii) $\beta=\gamma$  at $\lambda=0$;

(iv) the numerator of (\ref{eq:deff}) vanishes at $
\lambda=0,\epsilon=0, k^2\neq0$.

Consequently,
\begin{equation}
\label{eq:vanf1}
\stackrel{\lim}{\epsilon\to 0} f(\eta\cdot k
\neq0, k^2\neq0, \eta^2,\lambda=0,\epsilon)=0.
\end{equation}

Next, we treat the case $k^2=0$. Here:
(i) $i\xi_2=0$; $\xi_1=-i{\eta\cdot k\over\epsilon}$;

(ii) 
$\gamma=0;\ \beta={\lambda\epsilon^2\over{(\eta\cdot k)^2+i\epsilon\eta^2}}$;

(iii) the denominator
 of (\ref{eq:deff}) simplifies to
\begin{equation}
\label{eq:vanf2}
[-i\xi_1\kappa_2-1][i\xi_1\kappa_1-1]
= \xi_1^2\beta+2i\xi_1\sqrt{\gamma^2-\beta}+1
\end{equation}

Counting only the powers of $\epsilon$:
\begin{eqnarray}
\label{eq:vanf3}
& & [-i\xi_1\kappa_2-1][i\xi_1\kappa_1-1]\equiv O(\epsilon^0);\nonumber\\
& & \sqrt{\gamma^2-\beta}\equiv
 O(\epsilon\sqrt{\lambda}),\ \lambda\neq0;\nonumber\\
& & ln\biggl[{\kappa_2-\kappa_1\kappa_2
\over{\kappa_1-\kappa_1\kappa_2}}\biggr]\equiv O(\epsilon^0);
\nonumber\\
& & {1\over{1-i\xi_2+\xi_1^2+i\xi_2\xi_3}}
\equiv O(\epsilon^2),\ \eta\cdot k\neq 0
\nonumber\\
& & 
({\rm numerator}:)i\xi_1^2\beta-i\lambda+\sqrt{\gamma^2-\beta}i\xi_2\xi_3(1-\lambda)\equiv
O(\epsilon^0)
\end{eqnarray}
Then:
\begin{equation}
\label{eq:vanf4}
f(\eta\cdot k
\neq0, k^2=0, \eta^2,\lambda\neq0,\epsilon)=0(\epsilon)
\end{equation}
and we find:
\begin{equation}
\label{eq:vanf5}
\stackrel{\lim}{\epsilon\to 0} 
f(\eta\cdot k
\neq0, k^2=0, \eta^2,\lambda\neq0,\epsilon)=0.
\end{equation}
We thereafter set $\lambda=0$.

Finally, we shall deal
 with the subspace $k^2=\eta\cdot k=0$.
Here, we note  that $\eta^2\neq$ is necessary to begin with for
the procedure to be defined in the intermediate stages. Then:
\begin{eqnarray}
\label{eq:vanf6}
& & [-i\xi_1\kappa_2-1][i\xi_1\kappa_1-1]\equiv O(\epsilon^0);\nonumber\\
& & \gamma=0;\ \beta=
{\lambda\epsilon\over\eta^2}\nonumber\\
& & \kappa_1=-\kappa_2=\sqrt{-\beta}=
\sqrt{\gamma^2-\beta}\equiv 
O\sqrt{{\lambda\epsilon\over\eta^2}};\nonumber\\
& & ln\biggl[{\kappa_2-\kappa_1\kappa_2
\over{\kappa_1-\kappa_1\kappa_2}}\biggr]\equiv O(\epsilon^0);
\nonumber\\
& & {1\over{1-i\xi_2+\xi_1^2+i\xi_2\xi_3}}
\equiv O({\epsilon\over\eta^2}),\ \eta\cdot k\neq 0
\nonumber\\
& & ({\rm numerator}:)-i\lambda\equiv
O(\epsilon^0\lambda)
\end{eqnarray}
Thus,
\begin{equation}
\label{eq:vanf7}
f(\eta\cdot k
\neq0, k^2=0, \eta^2\neq0,\lambda\neq0,\epsilon)
= O(\sqrt{{\lambda\epsilon\over\eta^2}}).
\end{equation}
Thus we  may set $\lambda=0$. Then
\begin{equation}
\label{eq:vanf8}
f(\eta\cdot k
\neq0, k^2=0, \eta^2\neq0,\lambda=0,\epsilon)=0.
\end{equation}
We may then set $\eta^2=0$ at the end for LCG though
$\eta^2\neq0$ is needed for the procedure to be well defined.

One can show that a similar analysis holds good for the contribution of
(\ref{eq:secterm}) to the $\eta_\mu k_\nu$ terms in (\ref{eq:Result1}).

\section{}
\setcounter{equation}{0}
\seceqbb

In this appendix, we shall consider the
expression of  the first term in (\ref{eq:pres10}) involving
\begin{equation}
\label{eq:firterm1}
{1\over{a_1-\kappa_1}}ln\biggl[{\kappa_1-a_1\kappa_1\over{a_1-a_1\kappa_1}}
\biggr]
\end{equation}
which can be expanded in a Taylor series in powers of
${\kappa_1-a_1\over{a_1(1-\kappa_1)}}$ in the domain defined by:
\begin{equation}
\label{eq:firterm2}
|\kappa_1-a_1|<|a_1(1-\kappa_1)|.
\end{equation}
Further, the expression (\ref{eq:firterm1}) could be truncated  as:
\begin{equation}
\label{eq:firterm3}
-{1\over{a_1(1-\kappa_1)}}+O(a_1-\kappa_1)
\end{equation}
and higher order terms
neglected as $\epsilon\to 0$ if the expansion parameter
\begin{equation}
\label{eq:firterm4}
{\kappa_1-a_1\over{a_1(1-\kappa_1)}}\to0\ {\rm as}\ \epsilon\to0.
\end{equation}
In this appendix, we shall seek the domains of validity of (\ref{eq:firterm2}) and
(\ref{eq:firterm4}).

We divide out (\ref{eq:firterm2}) by $|\kappa_1||a_1|$ (valid 
except where $|\kappa_1|$ and $|a_1|$ vanish). The equation (\ref{eq:firterm2})
then translates as:
\begin{equation}
\label{eq:firterm5}
|{1\over\kappa_1}-{1\over a_1}|<|{1\over\kappa_1}-1|
\end{equation}
Defining ${1\over\kappa_1}-1\equiv Y$ 
and ${1\over a_1}-1\equiv-X=-i(\xi_1+\xi_2)$, (\ref{eq:firterm2})
reads:
\begin{equation}
\label{eq:firterm6}
|Y+X|<|Y|.
\end{equation}
This simplifies to
\begin{equation}
\label{eq:firterm7}
Re\biggl({Y\over X}\biggr)<-{1\over 2}.
\end{equation}
We recall:
\begin{eqnarray}
\label{eq:firterm8}
& & Y={1\over\kappa_1}-1={\kappa_2\over\beta}-1
={\gamma\over\beta}-1
-\sqrt{\biggl({\gamma\over\beta}\biggr)^2 - {1\over\beta}}\nonumber\\
& & ={-i\xi_2\lambda-
\sqrt{(1-i\xi_2)(-\xi_1^2-i\xi_2\xi_3)
-i\xi_2\lambda(1+\xi_1^2-i\xi_2+i\xi_2\xi_3)}
\over{1-i\xi_2(1-\lambda)}}
\end{eqnarray}

We analyze the condition (\ref{eq:firterm7}) at $\lambda=0$, which
is sufficient. The equation (\ref{eq:firterm7}) then  comes:
\begin{equation}
\label{eq:firterm9}
Im\biggl[\sqrt{{k^2+i\epsilon\over k^2}}
{\sqrt{-(\eta\cdot k)^2-i\epsilon\eta^2}\over{\eta\cdot k+\epsilon}}\biggr]
>{1\over2}.
\end{equation}
For $|\eta\cdot k|>>\epsilon$ and $|k^2|>>\epsilon$, it is  evident
that the left hand side is (using the convention for the square root
as given below (\ref{eq:a1kappa1}))
\begin{equation}
\label{eq:firterm10}
Im\biggl[{i\eta\cdot k\over{\eta\cdot k+\epsilon}}\biggr]
\approx1>{1\over 2}
\end{equation}
and thus (\ref{eq:firterm9}) is automatically satisfied
In fact, for $\eta\cdot k\to0^+$ also, (noting that 
the phase of  $\sqrt{-(\eta\cdot k)^2-i\epsilon\eta^2}$ varies 
continuously from ${\pi\over2}$ as $\eta\cdot k\to0^+$ from 
large values) (\ref{eq:firterm9}) reads, for $\eta^2\neq0$,
\begin{equation}
\label{eq:firterm11}
Im\biggl[{\sqrt{-i\epsilon\eta^2}\over\epsilon}\biggr]={1\over{\sqrt{2}}}
{\sqrt{|\eta^2|}\over\sqrt{\epsilon}}
\end{equation}
and is automatically $>{1\over2}$ for $\epsilon$ sufficiently small.
Also a shift of  variables $\eta\cdot  k+\epsilon=-\zeta$, will
allow one to concluded in a similar manner  that for
$\zeta\to0^+$(i.e. $\eta\cdot k+\epsilon\to0^-$)the condition
(\ref{eq:firterm9}) is fulfilled for  $\eta^2\neq0$. A careful analysis of
(\ref{eq:firterm9}) in fact shows that (\ref{eq:firterm9}) is
satisfied for $\eta^2\neq0$ and real $\eta\cdot k$
everywhere on  the real $\eta\cdot k$ 
axis except the interval $(-\epsilon,0)$. For
$\eta^2=0$, similarly, (\ref{eq:firterm9})
is valid over the real  $\eta\cdot k$ axis
except  the interval $(-\epsilon,\epsilon)$. This is valid for any
$k^2$.

Next, we analyze the condition (\ref{eq:firterm4}) required,
in addition, for the truncation of the expansion (\ref{eq:firterm3}). 
The condition (\ref{eq:firterm4}), in notations  below (\ref{eq:firterm5})
reads:
\begin{equation}
\label{eq:firterm12}
|1+{X\over Y}|\to0,\ {\rm as}\ \epsilon\to0.
\end{equation}
The condition (\ref{eq:firterm12})
can be analyzed at $\lambda=0$. It then reads:
\begin{equation}
\label{eq:firterm13}
|1-{i(\xi_1+\xi_2)\sqrt{1-i\xi_2}\over
{\sqrt{-\xi_1^2-i\xi_2\xi_3}}}|\to0,
\end{equation}
i.e.
\begin{equation}
\label{eq:firterm14}
|1-{i(\eta\cdot k)
\sqrt{{k^2\over{k^2+i\epsilon}}}\over{\sqrt{-(\eta\cdot k)^2-i\epsilon\eta^2}}}|
\to0 \ {\rm as}\ \epsilon\to0.
\end{equation}
For $k^2\neq0$, and fixed $(\eta\cdot k)^2\neq0$, we can , in fact, 
by taking $\epsilon$ arbitrarily
small, make the left-hand side of (\ref{eq:firterm13})
arbitrarily small. Thus, the conditions for  the validity of
the expansion of (\ref{eq:firterm1}) and
its truncation of (\ref{eq:firterm3}) hold valid for  any real
$\eta\cdot k$ outside the interval $(-\epsilon,0)$
for $\eta^2\neq0$ and $(-\epsilon,\epsilon)$ for $\eta^2=0$ provided
$k^2\neq0$. The above analysis
of (\ref{eq:firterm13}) however fails at $k^2=0$.

We find that for  $k^2=0$, while the condition  
(\ref{eq:firterm7}) is satisfied
in the domain mentioned, the condition (\ref{eq:firterm12}) is 
not satisfied.  So, in this domain, 
we may  not use the truncated expression for (\ref{eq:firterm1}).

\section{}
\setcounter{equation}{0}
\seceqcc

In this appendix, we shall carry out the check that the propagator
(\ref{eq:Result1}), via expression (\ref{eq:mainln}), in fact leads to the
naive propagator at $\epsilon=0$.

As noted in Appendix A, the contribution of the second
term in (\ref{eq:mainln}), vanishes at $\epsilon=0$. We shall now
evaluate the contribution of the first term
 in (\ref{eq:mainln}). 
As noted in (\ref{eq:smalln.k}), this term reduces to:
\begin{equation}
\label{eq:epsch1}
{1\over{a_1-\kappa_2}}{\alpha+a_1\over{a_1-\kappa_1}}
ln\biggl[{\kappa_1-a_1\kappa_2\over{a_1-a_1\kappa_1}}\biggr]
=-{(a_1+\alpha)(1-i\xi_1)^2(1+\xi_1^2)\over{2\xi_1^2}}
\end{equation} 
($\xi_1={\eta\cdot k\over k^2}$ at $\epsilon=0$).  Then the right-hand
side of (\ref{eq:Result1}) at $\epsilon=0$ reads:
\begin{eqnarray}
\label{eq:epsch2}
& & {i\over{(k^2)^2(1-i\xi_1)(1+\xi_1^2)}}\biggl(\biggl[{\xi_1+i\xi_3\over{1-i\xi_1}}
+[i\lambda-\xi_1(1-\lambda)]\biggr]
{(1-i\xi_1)^2(1+\xi_1^2)\over{2\xi_1^2}}k_\mu k_\nu
\nonumber\\
& & -(1+i\xi_1)\biggl[{1\over{1-i\xi_1}}-{1\over{1-i\xi_1}}\biggr]
{(1-i\xi_1)^2(1+\xi_1^2)\over{2\xi_1^2}}
\eta_\mu k_\nu +(k\to-k;\ \mu\leftrightarrow\nu)\biggr)\nonumber\\
& &
\end{eqnarray} 
A straightforward simplification then leads to the familiar result
(\ref{eq:eps=0}).

By using the identity
\begin{eqnarray}
\label{eq:identeps0}
& & \Sigma(\epsilon=0)=2i(\eta\cdot k)^2k^2{d\over{d\epsilon}} ln\Sigma_3
|_{\epsilon=0}=\nonumber\\
& & 2i(\eta\cdot k)^2k^2\biggl[-{(\lambda-1)
\over{\eta\cdot k}}+i{(\lambda-1)\over{2k^2}}
-{i\over{2(\eta\cdot k)^2}}(\lambda k^2+\eta^2)\biggr],
\end{eqnarray}
one can show that the $\epsilon=0$ limit of (\ref{eq:freslt1}) also gives
(\ref{eq:eps=0}).

\newpage
\begin{figure}[p]
\centerline{\mbox{\psfig{file=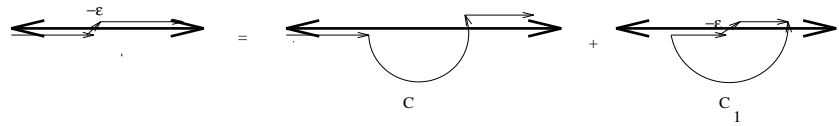,width=0.8\textwidth}}}
\caption{Contours $C$ and $C_1$}
\end{figure}

\begin{figure}[p]
\centerline{\mbox{\psfig{file=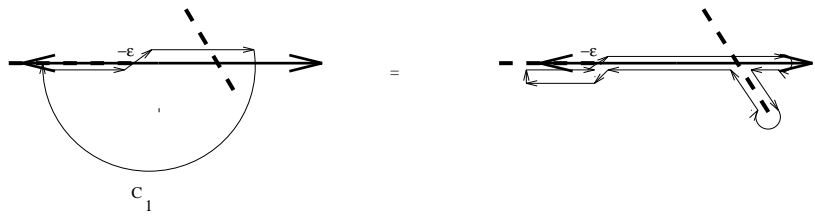,width=0.8\textwidth}}}
\caption{Distortion of contour $C_1$; 
the dashed lines indicate the branch cuts}
\end{figure}

\end{document}